\documentclass[printer]{aa} 

\bibpunct{(}{)}{;}{a}{}{,} 
\usepackage{natbib}
\usepackage{graphicx}
\usepackage{lscape}
\usepackage{rotating}
\usepackage{subfig}
\usepackage{hyperref}
\usepackage[T1]{fontenc}

\usepackage[varg]{txfonts}
\begin{document}

   \title{Evidence of a non-conservative mass transfer for XTE J0929-314}
%
%

\author{A. Marino\inst{\ref{inst1}} \and T. Di Salvo\inst{\ref{inst1}} \and A.\ F. Gambino\inst{\ref{inst1}} \and R. Iaria\inst{\ref{inst1}} \and L. Burderi\inst{\ref{inst2}} \and M. Matranga\inst{\ref{inst1}} \and A. Sanna\inst{\ref{inst2}} \and A. Riggio\inst{\ref{inst2}}}

\institute{Universit\`a degli Studi di
  Palermo, Dipartimento di Fisica e Chimica, via Archirafi 36 - 90123 Palermo, Italy\label{inst1}\\
  \email{alessio.marino@community.unipa.it} 
\and
           Universit\`a degli Studi di
  Cagliari, Dipartimento di Fisica, SP Monserrato-Sestu km 0.7, I-09042 Monserrato, Italy\label{inst2}
           }    

            
%


 %
  \abstract
   {In 1998 the first accreting millisecond pulsar, SAX J1808.4-3658, was discovered and to date 18 systems showing coherent, high frequency ($> 100$ Hz) pulsations in low mass X-ray binaries are known. Since their discovery, this class of sources has shown interesting and sometimes puzzling behaviours. In particular, apart from a few exceptions, they are all transient with very long X-ray quiescent periods implying a quite low averaged mass accretion rate onto the neutron star. Among these sources, XTE J0929-314 has been detected in outburst just once in about 15 years of continuous monitoring of the X-ray sky. } 
   {We aim to demonstrate that a conservative mass transfer in this system will result in an X-ray luminosity that is higher than the observed, long-term averaged X-ray luminosity. }
   {Under the hypothesis of a conservative mass transfer driven by gravitational radiation, as expected for this system given the short orbital period of about $ 43.6$ min and the low mass of the companion implied by the mass function derived from timing techniques, we calculate the expected mass transfer rate in this system and predict the long-term averaged X-ray luminosity. This is compared with the averaged, over 15 years, X-ray flux observed from the system, and a lower limit of the distance to the source is inferred.}
   {This distance is shown to be $> 7.4$ kpc in the direction of the Galactic anticentre, implying a large height, $> 1.8$ kpc, of the source with respect to the Galactic plane, placing the source in an empty region of the Galaxy. We suggest that the inferred value of the distance is unlikely. }
   {This problem can be solved if we hypothesize that the source is undergoing a non-conservative mass transfer, in which most of the mass transferred from the companion star is ejected from the system, probably because of the (rotating magnetic dipole) radiation pressure of the pulsar. If confirmed by future observations, this may be another piece of evidence that accreting millisecond pulsars experience a non-conservative mass transfer.}


\maketitle

\section{Introduction}
According to the recycling scenario \citep[see e.g.][for a review]{Bhattacharya1991}, that was recently directly confirmed \citep{Papitto2013}, there exists an evolutionary link between the so-called low-mass X-ray binaries (LMXBs) containing a neutron star and millisecond radio pulsars (MSPs). The first class of sources consists of old systems where a low magnetized ($\sim 10^8 - 10^9$ Gauss) neutron star accretes matter from a low-mass (usually less or of the order of a Solar mass, $M_\odot$) companion star. The weak magnetic field of the neutron star allows the matter to be accreted very close to the compact object; the accretion radius is indeed the magnetospheric radius (the radius at which the magnetic pressure due to the assumed dipolar magnetic field of the neutron star is balanced by the ram pressure of the accreting matter) which, for typical values of the magnetic field and the mass accretion rate, can be quite close (a few neutron star radii) to the compact object. In this situation, the neutron star can be accelerated by the accretion of matter and angular momentum from a (Keplerian) accretion disk to very short periods, in principle up to the limiting period (of the order of or below 1 ms), which depends on the mass-radius relation of the neutron star, and therefore on the equation of state of ultra-dense matter. At the end of the mass transfer phase, these systems will be observed as low magnetized, very fast spinning (millisecond) pulsars in a binary system with a very low-mass  companion star (if any); these systems are indeed observed in radio and form the class of MSPs.

This evolutionary scenario was confirmed by the discovery of millisecond coherent pulsations in LMXBs; this important discovery arrived in 1998, when coherent millisecond pulsations with a period of 2.5 ms where discovered in the transient LMXB SAX J1808.4-3658 \citep[][]{Wij1998}, thanks to the large effective area ($\sim 6000$ cm$^{-2}$ and high time resolution (up to $1 \ \mu$sec) of the Proportional Counter Array (PCA) on board the Rossi X-ray Timing Explorer (RXTE). SAX J1808.4-3658, otherwise a quite normal LMXB, belongs to a close binary system, $P_{orb} \simeq 2$ h \citep[][]{Chakrabarty1998}, and it is one of the few among the known accreting millisecond X-ray pulsars (in the following, we will refer to them as AMSPs) which has shown more than one X-ray outburst in the RXTE era. So far, we know of 18 accreting millisecond pulsars \citep[see][for a review]{Patruno_Watt2012}, the most recent one discovered by \citet{sanna2016}; all of them are X-ray transients in very compact systems (orbital period between 40 minutes and 4 hours), the fastest of which is IGR J00291+5934, with a spin period of $P_{spin} \simeq 1.7$ ms \citep{galloway2005}, and the slowest of which is XTE J0929-314, with a spin period of $P_{spin} \simeq 5.4$ ms \citep[][]{Galloway2002}.

XTE J0929-314 is a high-latitude source, the third discovered of the 18 AMSPs known to date. It was discovered by RXTE in 2002, when it showed an X-ray outburst which started on May 2 and lasted for about 53 days \citep[][]{Galloway2002}. One of the instruments on board RXTE, the All-Sky Monitor (ASM), witnessed the only two-month long outburst ever observed from this X-ray source. The signal, observed with the RXTE PCA, showed pulsations at a frequency of $185$ Hz (corresponding to a spin period of $5.4$ ms), revealing the presence of a millisecond pulsar. By Doppler modulations of the pulsating signal, the system was later recognized as an ultracompact binary. Timing analysis clearly pointed out that the orbital period of the binary was $43.6$ min. The last piece of the puzzle was the mass of the companion, extracted by the mass function of the system, $f_x=2.7 \times 10^{-7}\, M_{\odot}$, one of the smallest values ever found for such astronomical objects; under the hypothesis that the primary is a neutron star (NS) and has a mass of $1.4 \ M_{\odot}$, the value for the minimum mass of the secondary is $0.0083 M_{\odot}$. 
The source also showed a steady spin-down during the outburst, with $\dot \nu \sim -9.2(4) \times 10^{-14}$ Hz/s \citep{Galloway2002, DiSalvo2008AIPC}, possibly indicating a relatively high magnetic field of the neutron star.

In this paper, we show that the long-term mass accretion rate observed in XTE J0929-314 is not compatible with the mass accretion rate predicted by a conservative evolution of the system driven by gravitational radiation (GR), unless the source is placed at a large distance from our Solar system of $\ge 7.4$ kpc. At this distance, the height of the source above the Galactic plane would be of $\ge 1.8$ kpc towards the external part of the Milky Way, placing the source in an empty region of the Galaxy. We argue, therefore, that this source may experience a non-conservative mass transfer, as has been proposed for the case of SAX J1808.4-3658  \citep{DiSalvo2008, Burderi2009} and SAX J1748.9-2021 (Sanna et al. 2016) in order to explain the strong orbital expansion observed in these sources. 

\section{Data analysis and results}
The aim of this paper is to calculate the averaged mass accretion rate observed for this source in order to compare it with the expected averaged mass accretion rate for a conservative orbital evolution. \citet{Galloway2002} gave the first estimate of the mass transfer rate in this system, concluding that the distance to the source should be higher than 5 kpc assuming an outburst recurrence time higher than 6.5 yr. Considering that today we can state that the outburst recurrence time is higher than 15 yr, we can scale the lower limit to the distance by a factor $\sim 1.5$ showing that the source is placed at more than 7.6 kpc. In the following we critically repeat this argument in order to obtain a stringent lower limit on the source distance.

We have therefore started from the only outburst observed from this source in 2002. After XTE J0929-314 was identified by  the ASM, a series of pointed RXTE observations of the source was performed during 2002 from May 2 to June 24 (In Modified Julian Days from 52,396 to 52,449). The RXTE/PCA and ASM light curves of the source have been published in \cite{Galloway2002}.  
We can estimate the observed $2-10$ keV flux during the outburst by analyzing the light curves reported by these authors.
The area subtended by the curve ($2-10$ keV observed fluence) gives us the value of the flux integrated over the outburst duration, that corresponds to the observed fluence during the outburst, that is,\ the total energy per unit area emitted during the outburst as received by the detector. Approximating the shape of the curve with a triangle with height equal to the maximum of the curve and base equal to the total duration of the outburst, it is easy to find this fluence. In Table \ref{tab:Outburst} we report the data used to estimate the value of the $2-10$ keV fluence, together with the estimated 2–60 keV bolometric fluence over the entire outburst as reported in \citet{Galloway2002}.
\linebreak

\begin{table}[h!]
\centering
\begin{tabular}{ l  l }
\hline \hline
{Parameter} & {Value}\\
\hline
{Outburst start (MJDs)} & {52380}\\
{Outburst end (MJDs)}&{52445}\\
{Outburst duration (days)}&{65}\\
{Peak flux (erg $\cdot$ cm$^{-2}$)} & {$(7 \pm 1) \times 10^{-10}$}\\
{Observed fluence (erg $\cdot$ cm$^{-2}$)} & {$(2.0 \pm 0.3) \times 10^{-3}$} \\
{Bolometric fluence (erg $\cdot$ cm$^{-2}$)} & {$4.2 \times 10^{-3}$} \\
\hline
\end{tabular}
\caption{Details of the 2002 outburst of XTE J0929-314. The peak flux and the obs. fluence are as observed by the PCA in the 2-10 keV energy range. The bol.\ fluence is calculated in the $2-60$ keV energy range, from \citet{Galloway2002}.}
\label{tab:Outburst}
\end{table}

The observed fluence in the $2-10$ keV range is about a factor 2.1 smaller than the bolometric fluence ($2-60$ keV), in agreement with the bolometric correction factor of $2.34 \pm 0.12$, calculated as the mean ratio of the integrated $2-60$ keV to $2-10$ keV flux given by \citet{Galloway2002}. We will therefore use in the following the bolometric fluence reported in Table\ \ref{tab:Outburst}, and will correct the observed peak flux with the bolometric correction factor, which gives a bolometric peak flux of $\sim 1.6 \times 10^{-9}$ erg cm$^{-2}$ s$^{-1}$. 
To get the total energy emitted during the outburst we have to multiply this energy per unit area by a factor of $4\pi d^2$ \footnote{The surface over which this energy has been spread.}, assuming an isotropic emission from the source, where $d$ is the distance to the source (unknown at the moment). 
In this way we can calculate the total amount of energy released in the $2-60$ keV band during the outburst, $E_{bol}(d)$, that is $\sim 5 \times 10^{41}\, d^2_{kpc} erg$, where $d_{kpc}$ is the distance in units of 1 kpc. 
From this energy, we can calculate the averaged luminosity of the source and therefore the averaged mass accretion rate, which can be compared with the averaged mass transfer rate expected in the case of a conservative orbital evolution driven by GR.

Because XTE J0929-314 has shown only one outburst since its discovery, we can easily assume that the whole energy emitted by the system in the twenty years of monitoring\footnote{In 1996, the ASM onboard RXTE and the Wide Field Cameras (WFC) onboard BeppoSAX started a continuous monitoring of the X-ray sky. This is today continued by MAXI onboard the International Space Station, the Swift/BAT (Burst Alert Telescope) hard X-ray monitor, INTEGRAL, and the Gamma-ray Busrt Monitor (GBM) onboard Fermi.} is equal to the amount emitted during the outburst, since we can safely neglect the energy emitted during quiescent periods when the source luminosity is several (up to six) orders of magnitude below the outburst luminosity. Under this approximation, an estimate of the averaged X-ray luminosity is given just by  $L_{tot} = E_{tot}(d) / 20$ years, resulting in an averaged X-ray luminosity of 
$L_{tot} \sim 7.9 \times 10^{32}\, d^2_{kpc}$ erg/s in the $2-60$ keV band.
More conservatively, since we cannot exclude that the source went into outburst about 30 years ago, we can safely state that the recurrence time for the outbursts in this system is higher than (or of the order of) 15 years. In this case, the estimated bolometric luminosity from the source is $\le 10^{33}\, d^2_{kpc}$ erg/s. 
In the following we compare this value with the averaged X-ray luminosity expected from conservative mass transfer driven by GR in order to get an estimate of the distance to the source.

We adopted a theoretical model for the secular evolution of our system based on the hypothesis that mass transfer between the stars is conservative: if a quantity $m$ is lost from the secondary, the very same quantity of mass is accreted on the primary. In this model we consider an LMXB in which the secondary star is fully convective and is characterized by a mass that is less than $0.3 M_{\odot}$ \citep[see e.g.][]{NelsonRapp2003}, a condition which is compatible with many AMSPs. This approximation seems to be realistic because of the small orbital periods of this class of binaries (from less than a hour to a few hours); in this scenario the equation \citep[see e.g.][]{Verbunt1993}
\begin{equation}\label{eq:1}
P_{orb} \approx 8.9 \left(\frac{R_2}{R_{\odot}}\right)^{3/2} \left(\frac{M_{\odot}}{M_2}\right)^{1/2}\, hr
\end{equation}
tells us that for $P_{orb} \leq 3$ hr the secondary star mass is $M_2 \leq 0.3 M_{\odot}$. Moreover, according to the small mass function of XTE J0929-314, a mass greater than $0.25 M_{\odot}$ would require an inclination smaller than $2^\circ$, which is unlikely. The companion mass in this system is less than $0.03\, M_\odot$ (at 95\% confidence) for a uniform a priori distribution in $\cos i$ and for a neutron star mass of $2\, M_\odot$, close to the maximum mass for a neutron star \citep{Galloway2002}. 

The mass-radius relation for this category of stars is $n=-1/3$. This value for $n$ is the proper one for degenerate stars, but it is possible to show that it is valid for stars with masses $M_2 \leq 0.3 5M_{\odot}$ too (Nelson \& Rappaport, 2003). Under this mass-limit, in fact, a star becomes fully convective and its thermal timescale becomes too large, so that the star is out of thermal equilibrium. 
It then reacts on an adiabatic timescale with index $ n=-1/3$, making the previous assumption good in any case \citep[see also][who specifically address, for the first time to our knowledge, how the mass transfer rate depends on the adiabatic response of the mass-losing companion]{1982ApJ...254..616R}. Since in a fully convective star the magnetic braking is inhibited for the companion star, GR is the only available channel for the system to lose angular momentum. We also assume the contact condition: $\dot{R}_{L2}/R_{L2}=\dot{R_2}/R_2$, that guarantees the stability of mass transfer. Using the equation \citep[see][for a derivation]{Verbunt1993}
\begin{equation}\label{eq:Evo}
\frac{\dot{R}_{L2}}{R_{L2}}=2\frac{\dot{L}}{L}-2\frac{\dot{M_2}}{M_2}\left(\frac{5}{6}-\frac{M_2}{M_1}\right),
\end{equation}
along with the Paczynski's formula for the radius of the Roche lobe (in this scenario, assuming $M_1=1.4 M_{\odot}$, the ratio $q = m_2/m_1$ of the masses is certainly lower than $0.8$) and the assumptions previously discussed, we found the following equation for $\dot{M}$ and $\dot{P}$ in the framework of a conservative mass-exchange:  
\begin{equation}\label{eq:Mdot}
\dot{M} = -\dot{M_2} = 4.03 \times 10^{-9}\left[\frac{q^2}{\left(- \frac{2}{3}+2g\right)\left(1+q\right)^{1/3}}\right]m_1^{8/3}P_{2h}^{-8/3}\, M_{\odot} yr^{-1},
\end{equation}
\begin{equation}\label{eq:Pdot}
\dot{P}_{orb}=9.2 \times 10^{-13}\left[\frac{q}{\left(- \frac{2}{3}+2g\right)\left(1+q\right)^{1/3}}\right]m_1^{5/3}P_{2h}^{-5/3}\, s s^{-1},
\end{equation}
where $g = 1-q$ for a conservative mass transfer and the value $n = -1/3$ has been already used in the equations \citep[see also][]{DiSalvo2008, Burderi2009}.
Equation \ref{eq:Mdot} for $\dot{M}$ predicts the X-ray luminosity $L_X$ of the system; in fact, for neutron star binaries powered by accretion, X-ray luminosity is essentially $L=\eta GM\dot{M}/R$, where $\eta$ is a parameter of the order of unity accounting for any uncertainty in the conversion from mass accretion rate to X-ray luminosity, including possible relativistic corrections which are anyway of the order of $10\%$ at most.
Assuming a $1.4\, M_\odot$ neutron star, a value of $12$ km for its radius - a standard value for neutron stars based on recent observational and theoretical inferences \citep[see e.g.][]{Ozel2016}, a companion mass of $0.01\, M_\odot$ - corresponding to a mass ratio of $q = 0.007$, and the orbital period of the system, we obtain: 
\begin{equation}\label{eq:LX}
L_X = 5.5 \times 10^{34} \ \eta\ {\rm erg/s}
\end{equation}
The predicted mass transfer rate given by Eq.\ \ref{eq:Mdot} has a strong dependence on the mass ratio $q$ of the system, as $q^2$, and increases by increasing the mass ratio of the system. However, we have used for the companion mass a very low value of $0.01\, M_\odot$, very close to the minimum companion mass (corresponding to the maximum inclination angle of $90^\circ$) of $0.008\, M_\odot$. Assuming a more realistic value of the inclination angle, that should be smaller than $75^\circ$, since no dips or eclipses are observed in this system, we get a companion star mass larger than $\sim 0.01\, M_\odot$ for a $1.4\, M_\odot$ neutron star. Therefore, in this expression we have used the minimum realistic value for the companion star mass.  
We note also that $1.4\, M_\odot$ is a good lower limit for the mass of a recycled neutron star. Indeed, the minimum mass for a neutron star is $1.1-1.2\, M_\odot$, and the smallest measured neutron star mass to date is $1.174 \pm 0.004 \, M_\odot$ \citep[see][]{Martinez2015} observed in the 4.07-day binary pulsar J0453+1559 under the hypothesis that it is a double neutron star system, as seems to be indicated by the orbital eccentricity of the system ($e \simeq 0.11$). 
According to \citet{Ozel2012}, who investigated the distribution of neutron star masses in different populations of binaries employing Bayesian statistical techniques, the distribution of neutron star masses in non-recycled eclipsing high-mass binaries, as well as the distribution of slow pulsars, which are all believed to be near their birth masses, has a mean of $1.28\, M_\odot$ and a dispersion of $0.24\, M_\odot$, which are consistent with expectations for neutron star formation in core-collapse supernovae. On the other hand, the mass distribution of neutron stars that have been recycled has a mean of $1.48 \, M_\odot$ and a dispersion of $0.2\, M_\odot$, consistent with the expectation that they have experienced extended mass accretion episodes. Since XTE J0929-314 is a millisecond spinning, recycled neutron star, probably at the end of its life as a LMXB, we think that $1.4\, M_\odot$ is a reasonable lower limit for its neutron star mass.
As regards the radius adopted for the neutron star, high quality datasets from X-ray satellites, as well as significant progress in theoretical modelling, led to considerable progress in the measurements, placing them in the $9.9 - 11.2$ km range and shrinking their uncertainties due to a better understanding of the sources of systematic errors \citep[see][and references therein]{Ozel2016}. Hence, the adopted value of 12 km can be safely considered a reasonable upper limit to the neutron star radius.

Under the hypothesis of conservative mass transfer, we can impose $L_{tot}=L_X$, and we can solve this equation to find the distance to the source. Considering the averaged X-ray luminosity over the last 20 years of $L_{tot} \sim 7.9 \times 10^{32}\, d^2_{kpc}$ erg/s in the $2-60$ keV band, we get a distance to the source of $\sim 8.4 \eta^{1/2}$ kpc. Assuming from now on $\eta \simeq 1$, we also get a peak luminosity during the outburst of $\sim 5.9 \times 10^{36}$ erg/s, without considering the bolometric correction factor, and $\sim 1.3 \times 10^{37}$ erg/s including the bolometric correction factor. With the more conservative assumption of a recurrence time of 15 years for the outburst in this source, we get $L_{tot} \le 10^{33}\, d^2_{kpc}$ erg/s, corresponding to a distance to the source of $d \ge 7.4$ kpc. 
At a distance larger than 7.4 kpc, the flux at the peak of the 2002 outburst of the source should be $L_{peak} \ge  10^{37}$ erg/s, taking into account the bolometric correction factor; this peak luminosity is higher than the averaged X-ray luminosity at the peak of the outburst in AMSPs (usually a few $10^{36}$ erg/s), but still acceptable. 

\section{Discussion}
\label{sec:discussion}

In the previous section we have evaluated the observed averaged X-ray luminosity of the AMSP XTE J0929-314 as a function of its distance and we have compared this estimate with the lower limit on the mass accretion rate in the system, under the hypothesis of a conservative mass transfer driven by GR. In this way we have obtained a lower limit to the distance to the source of 7.4 kpc. This value is compatible to the lower limit to the source distance given by \citet{Galloway2002}, with similar considerations when extrapolated to take into account the updated recurrence time of the outburst of about 15 yr ($d \ge 7.4$ kpc).

XTE J0929-314 has a high Galactic latitude $b=14.2^\circ$, towards the Galactic anticentre, so its height from the equatorial plane of the Galaxy should be $h = d \sin{b} \ge 1.8$ kpc.
These results are summarized in Table \ref{tab:Source}.

\begin{table}[h!]
\centering
\begin{tabular}{ l  l }
\hline \hline
{Parameter} & {Value}\\
\hline
{Theoretical luminosity (erg/s)} & {$5.5 \times 10^{34}$}\\
{Distance on the equatorial plane (kpc)} & {$\ge 7.4$}\\
{Height from the equatorial plane (kpc)} & {$\ge 1.8$}\\
\hline
\end{tabular}
\caption{Theoretical luminosity for a conservative mass transfer and constraints 
on the location of XTE J0929-314 in our Galaxy.}
\label{tab:Source}
\end{table}

\begin{figure}[t!]
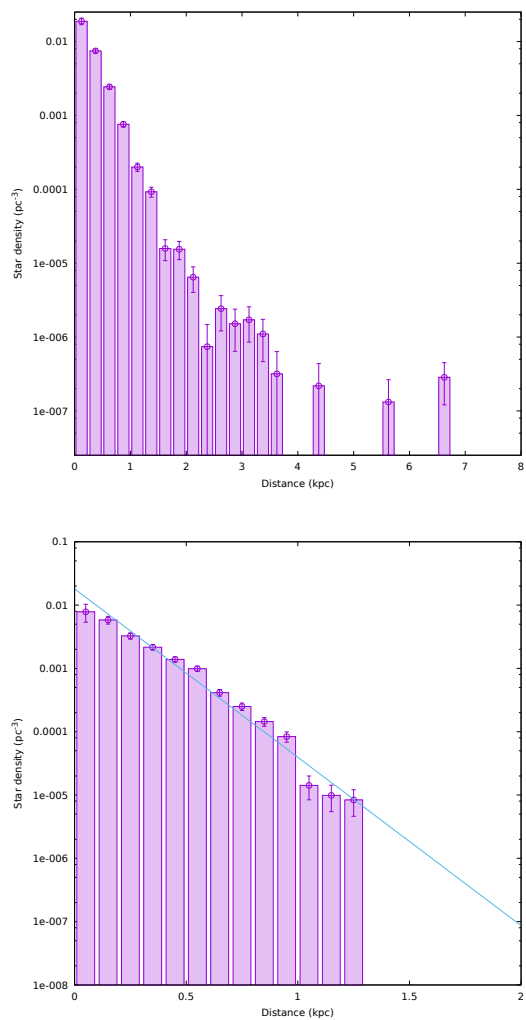

\centering
\includegraphics[scale=0.7]{gaia_xte_j0929_2deg.pdf}
\includegraphics[scale=0.7]{Gaia2.pdf}
\caption{Stellar density as a function of the distance along the direction of XTE J0929-314 within $2^\circ$, obtained using the GAIA DR1 catalogue which includes precise measurements of the stellar parallax. Upper panel: all the stars for which a parallax has been measured. Bottom panel: stars with precise parallax measurements (with $d / \delta d \ge 3$); the solid line on top of the data is the best-fit to the stellar density as a function of the distance using an exponential decay law (see text for more details).}
\label{fig:Gaia}
\end{figure}

The height of the source over the Galactic plane obtained following this procedure is about six times the size of the thin stellar disk of our Galaxy \citep[assuming a scale height of the thin stellar disk of $\sim 0.3$ kpc, see e.g.][]{RixBovy2013}.
No globular clusters are known to be present in proximity of this position. It seems unlikely to have an X-ray binary at such a high height with respect to the Galactic plane. 
To study the stellar distribution in the milky way along the direction of the source, we take advantage of the very recently published first astrometric catalogue of the mission GAIA, \citep[GAIA DR1, see][]{Gaia.etal:2016b, Gaia.etal:2016a, Lindegren.etal:2016} with measurements of the parallax of over one billion stars in our Galaxy. We extracted from GAIA DR1 all the stars with an angular distance from XTE J0929-314 less than $2^\circ$ (Fig. \ref{fig:Gaia}, upper panel). From this figure it is evident that the density of stars above $\sim 4$ kpc drops significantly. However, the most distant stars in this figure have large uncertainties on the measured parallax (this uncertainty will probably be improved with time, when GAIA will acquire a significant number of passages on each of these objects). We have also selected those stars for which the distance is well constrained ($d / \delta d > 3$), which are shown in Figure \ref{fig:Gaia} (bottom panel). To show the reduction of the stellar density as a function of the distance in the direction of the source we have fitted these data; the stellar density is well fitted by an exponential law, $\rho(d) = \rho_0 \exp(-d/d_0)$, where $\rho_0$ is the density at the Earth position and $d_0$ is the scale length. In this case, $d_0 \simeq 0.2$ kpc. In this way, we have obtained the stellar density $\rho(d)$ as a function of the distance $d$, which is also shown in Figure \ref{fig:Gaia} (bottom panel, solid line on top of the data). 
From the figure it is clear that the density of stars in the direction of the source significantly drops above 2 kpc, and in particular a very small number of stars are present above 7 kpc. We argue therefore that it is unlikely that the source is located at a distance higher than 4 kpc.

In other words it is unlikely that the source is experiencing a conservative mass transfer. We note that the longer the source remains in X-ray quiescence, the larger will be the discrepancy between the mass accretion rate predicted for a conservative mass transfer and the averaged luminosity of the source, and the larger will be the inferred distance to the source. Therefore, the distance we infer here should be considered as a lower limit to the distance to the source under the hypothesis of a conservative mass transfer in the system.

On the other hand, if the mass released from the secondary star is, at least in part, expelled from the binary system in a non-conservative mass transfer, the accretion luminosity would be smaller than what is obtained from a conservative mass transfer, affecting our determination of the distance to the source.
We can formulate an equation similar to Eq. \ref{eq:Evo} adapted for the non-conservative case, to describe the case of a non-conservative mass transfer in which part of the transferred matter is ejected from the system
\begin{equation}\label{eq:EvoNC}
\frac{\dot{P}_{orb}}{P_{orb}}=3\left\lbrace\frac{\dot{J}}{J_{orb}}-\frac{\dot{M_2}}{M_2}\left[1-\beta q - \frac{\left(1-\beta\right)\left(\alpha+\frac{q}{3}\right)}{1+q}\right]\right\rbrace,
\end{equation}
where $\beta$ is the fraction of the mass transferred from the secondary that is accreted onto the NS.
However, without any information about the value of $\dot{M_2}$, we lack the tools necessary to solve this equation at present. We could, however, get out of this impasse in the future if we were able to obtain an independent measure of the distance to the source or of the rate $\dot{P}_{orb}$ at which the orbital period changes in time. 

For the moment, we can at least argue a reasonable distance for the source as the one that makes the theoretical orbital period derivative, $\dot{P}_{orb}$, comparable to the rates measured for other AMSPs. In order to make this estimate, we have to make some necessary assumptions: (i) the amount of non-accreted matter is ejected from the inner Lagrangian point and (ii) $\beta$ is equal to the ratio between the outburst duration and the quiescence time (so, in the present case, we will assume $\beta=(65 / 5500) \approx 0.01$). The predicted value of the orbital period derivative can be calculated from Eq.\ \ref{eq:EvoNC} that can be re-arranged (for angular momentum losses via mass-ejection and GR only) to obtain Eq.\ (2) of \citet{Burderi2009}.
In this equation, $\alpha$ is the specific angular momentum of the mass lost from the system in units of the specific angular momentum of the companion that, under hypothesis (i), is $\sim 0.7$, and
$\dot{m}_{-9}$ is the mass transfer rate in units of $10^{-9}\, M_\odot/yr$. 
To estimate this parameter we used the mass transfer rate obtained setting the luminosity $L_{tot}$ of the system found above as equal to the accretion luminosity $L=GM\dot{M}/R$ and solving for $\dot{M}(d^2_{kpc})$ to find the rate of mass accreted. Considering that only a fraction $\beta$ of the mass transferred by the secondary is accreted onto the NS, we have divided this value for $\beta$ to determine the mass transfer rate. Solving the equation we calculate the distance 
for which the orbital period derivative attains a value of $10^{-12}$ s/s, of the same order of magnitude of the orbital period derivative measured for SAX J1808.4-3658 \citep[][]{DiSalvo2008, Burderi2009}. This distance is approximately $2.8 \ kpc$ and is definitely less unlikely than the one found in the conservative mass-transfer scenario.

We can ask ourselves why the system should experience a non conservative mass transfer. One of the theories developed to explain the origin of such a non-conservative evolution for AMSPs is the radio-ejection model, discussed extensively by \cite{Burderi2001}. The basic idea is that a fraction of the transferred matter in the disk could be swept out by radiative pressure of the pulsar. We summarize  here the fundamental points of the model. Accretion disk stability in this case is granted by the balance between essentially two pressures: the ram pressure of the disk, directly related to transfer rate $\dot{M_2}$ and directed towards the neutron star, and the magnetic or radiation pressure, due to the emission of the pulsar, directed outwards. 
Once the system experiences a reduction of the mass accretion rate (caused by a disk instability model or other mechanisms) such that the inner disk radius is pushed beyond the light cylinder radius, the radiation pressure from the pulsar (emitting as a magnetic-dipole rotator) may be able to sweep away from the system the matter in the disk and expel the transferred matter directly at the inner Lagrangian point, giving rise to a non-conservative mass transfer.

This model has also been used to explain the large orbital period derivative observed in SAX J1808.4-3658 \citep[][]{DiSalvo2008, Burderi2009}. For this source, the measured orbital period derivative is $\dot P_{orb} \sim 4 \times 10^{-12}$ s/s.  This derivative implies orbital expansion, from which we infer that the mass-radius index for the secondary should be $n < 1/3$ (we assume $n = -1/3$ for the mass-radius index of the secondary, in the reasonable hypothesis that the secondary star is a fully convective star out of thermal equilibrium and responds adiabatically to the mass transfer).
This derivative is about a factor of 70 higher than the orbital derivative expected for conservative mass transfer given the low averaged mass accretion rate onto the neutron star; since SAX J1808.4-365 accretes for about 30 d every $2-4$ yr, we have estimated an order of magnitude for the averaged X-ray luminosity from the source that is $L_X \sim 4 \times 10^{34}$ erg/s. 

A non-conservative mass transfer can explain the large orbital period derivative if we assume a mass transfer of $\sim 10^{-9}\, M_\odot$/yr, and that this matter is expelled from the system with the specific angular momentum at the inner Lagrangian point. 
However, the large orbital period derivative observed in SAX J1808.4-3658 has been alternatively interpreted as the effect of short-term angular momentum exchange between the mass donor and the orbit,\ resulting from gravitational quadrupole coupling due to variations in the oblateness of the companion star \citep{Hartman2009}. This interpretation has been strengthened by the observation of an accelerated orbital period expansion, at a rate of $\ddot{P}_{orb} \simeq 1.6 \times 10^{-20}\,s\,s^{-2}$, observed including the timing results during the 2011 outburst, and by extending the long-term measurements of the orbital evolution over a baseline of 13 years \citep{Patruno2012_b}. Therefore, it is still unclear whether the orbital variation observed in SAX J1808.4-3658 can be ascribed to a strong outflow of matter from the system, caused by a non-conservative mass transfer induced by the radiation pressure of the pulsar which is able to expel most of the transferred mass from the inner Lagrangian point, or to a short-term variability induced by tidal dissipation and magnetic activity in the companion, which is required to be at least partially non-degenerate, convective, and magnetically active \citep[][]{Applegate}, although in this case the orbital period derivative should change sign on a 10 yr timescale. 

The arguments developed here for XTE J9029-314 add another independent piece of evidence that AMSPs experience a non-conservative mass transfer induced by so-called radio ejection. It explains how the evolution of AMSP systems could go on evenly in a regime of non-conservative mass transfer and, at the same time, it is still consistent with all the consolidated theory about LMXBs; in particular, the radio ejection model could be able to fill the gap and the incongruences between a model of orbital evolution in which the total amount of mass is accreted onto the neutron star and the puzzling experimental results obtained for XTE J0929-314 and other AMSPs.


\section*{Acknowledgements}

We thank the anonymous referee for his/her detailed and useful report 
which helped us to improve our manuscript.
This work has made use of data from the European Space Agency (ESA)
mission {\it Gaia} (\url{http://www.cosmos.esa.int/gaia}), processed by
the {\it Gaia} Data Processing and Analysis Consortium (DPAC,
\url{http://www.cosmos.esa.int/web/gaia/dpac/consortium}). Funding
for the DPAC has been provided by national institutions, in particular
the institutions participating in the {\it Gaia} Multilateral Agreement.
We acknowledge financial contributions from the agreement ASI-INAF I/037/12/0. 
This work was partially supported by the Regione Autonoma della
Sardegna through POR-FSE Sardegna 2007-2013, L.R. 7/2007,
Progetti di Ricerca di Base e Orientata, Project N. CRP-60529.
AR acknowledges the Sardinia Regional Government
for its financial support (P.O.R. Sardegna F.S.E. Operational
Programme of the Autonomous Region of Sardinia, European Social
Fund 2007-2013 - Axis IV Human Resources, Objective l.3, Line of
Activity l.3.1.).

\bibliographystyle{aa} 
\bibliography{biblio}

\begin{thebibliography}{25}
\expandafter\ifx\csname natexlab\endcsname\relax\def\natexlab#1{#1}\fi

\bibitem[{{Applegate} \& {Shaham}(1994)}]{Applegate}
{Applegate}, J.~H. \& {Shaham}, J. 1994, \apj, 436, 312

\bibitem[{{Bhattacharya} \& {van den Heuvel}(1991)}]{Bhattacharya1991}
{Bhattacharya}, D. \& {van den Heuvel}, E.~P.~J. 1991, \physrep, 203, 1

\bibitem[{{Burderi} {et~al.}(2001){Burderi}, {Possenti}, {D'Antona}, {Di
  Salvo}, {Burgay}, {Stella}, {Menna}, {Iaria}, {Campana}, \&
  {d'Amico}}]{Burderi2001}
{Burderi}, L., {Possenti}, A., {D'Antona}, F., {et~al.} 2001, \apjl, 560, L71

\bibitem[{{Burderi} {et~al.}(2009){Burderi}, {Riggio}, {di Salvo}, {Papitto},
  {Menna}, {D'A{\`i}}, \& {Iaria}}]{Burderi2009}
{Burderi}, L., {Riggio}, A., {di Salvo}, T., {et~al.} 2009, \aap, 496, L17

\bibitem[{{Chakrabarty} \& {Morgan}(1998)}]{Chakrabarty1998}
{Chakrabarty}, D. \& {Morgan}, E.~H. 1998, \nat, 394, 346

\bibitem[{{Di Salvo} {et~al.}(2008{\natexlab{a}}){Di Salvo}, {Burderi},
  {Riggio}, {Papitto}, \& {Menna}}]{DiSalvo2008}
{Di Salvo}, T., {Burderi}, L., {Riggio}, A., {Papitto}, A., \& {Menna}, M.~T.
  2008{\natexlab{a}}, \mnras, 389, 1851

\bibitem[{{Di Salvo} {et~al.}(2008{\natexlab{b}}){Di Salvo}, {Burderi},
  {Riggio}, {Papitto}, \& {Menna}}]{DiSalvo2008AIPC}
{Di Salvo}, T., {Burderi}, L., {Riggio}, A., {Papitto}, A., \& {Menna}, M.~T.
  2008{\natexlab{b}}, in American Institute of Physics Conference Series, Vol.
  1054, American Institute of Physics Conference Series, ed. M.~{Axelsson},
  173--182

\bibitem[{{Gaia Collaboration} {et~al.}(2016{\natexlab{a}}){Gaia
  Collaboration}, {Brown}, {Vallenari}, {Prusti}, {de Bruijne}, {Mignard},
  {Drimmel}, {Babusiaux}, {Bailer-Jones}, {Bastian}, \&
  et~al.}]{Gaia.etal:2016b}
{Gaia Collaboration}, {Brown}, A.~G.~A., {Vallenari}, A., {et~al.}
  2016{\natexlab{a}}, \aap, 595, A2

\bibitem[{{Gaia Collaboration} {et~al.}(2016{\natexlab{b}}){Gaia
  Collaboration}, {Prusti}, {de Bruijne}, {Brown}, {Vallenari}, {Babusiaux},
  {Bailer-Jones}, {Bastian}, {Biermann}, {Evans}, \& et~al.}]{Gaia.etal:2016a}
{Gaia Collaboration}, {Prusti}, T., {de Bruijne}, J.~H.~J., {et~al.}
  2016{\natexlab{b}}, \aap, 595, A1

\bibitem[{{Galloway} {et~al.}(2002){Galloway}, {Chakrabarty}, {Morgan}, \&
  {Remillard}}]{Galloway2002}
{Galloway}, D.~K., {Chakrabarty}, D., {Morgan}, E.~H., \& {Remillard}, R.~A.
  2002, \apjl, 576, L137

\bibitem[{{Galloway} {et~al.}(2005){Galloway}, {Markwardt}, {Morgan},
  {Chakrabarty}, \& {Strohmayer}}]{galloway2005}
{Galloway}, D.~K., {Markwardt}, C.~B., {Morgan}, E.~H., {Chakrabarty}, D., \&
  {Strohmayer}, T.~E. 2005, \apjl, 622, L45

\bibitem[{{Hartman} {et~al.}(2009){Hartman}, {Patruno}, {Chakrabarty},
  {Markwardt}, {Morgan}, {van der Klis}, \& {Wijnands}}]{Hartman2009}
{Hartman}, J.~M., {Patruno}, A., {Chakrabarty}, D., {et~al.} 2009, \apj, 702,
  1673

\bibitem[{{Lindegren} {et~al.}(2016){Lindegren}, {Lammers}, {Bastian},
  {Hern{\'a}ndez}, {Klioner}, {Hobbs}, {Bombrun}, {Michalik}, {Ramos-Lerate},
  {Butkevich}, {Comoretto}, {Joliet}, {Holl}, {Hutton}, {Parsons},
  {Steidelm{\"u}ller}, {Abbas}, {Altmann}, {Andrei}, {Anton}, {Bach},
  {Barache}, {Becciani}, {Berthier}, {Bianchi}, {Biermann}, {Bouquillon},
  {Bourda}, {Br{\"u}semeister}, {Bucciarelli}, {Busonero}, {Carlucci},
  {Casta{\~n}eda}, {Charlot}, {Clotet}, {Crosta}, {Davidson}, {de Felice},
  {Drimmel}, {Fabricius}, {Fienga}, {Figueras}, {Fraile}, {Gai}, {Garralda},
  {Geyer}, {Gonz{\'a}lez-Vidal}, {Guerra}, {Hambly}, {Hauser}, {Jordan},
  {Lattanzi}, {Lenhardt}, {Liao}, {L{\"o}ffler}, {McMillan}, {Mignard}, {Mora},
  {Morbidelli}, {Portell}, {Riva}, {Sarasso}, {Serraller}, {Siddiqui}, {Smart},
  {Spagna}, {Stampa}, {Steele}, {Taris}, {Torra}, {van Reeven}, {Vecchiato},
  {Zschocke}, {de Bruijne}, {Gracia}, {Raison}, {Lister}, {Marchant},
  {Messineo}, {Soffel}, {Osorio}, {de Torres}, \&
  {O'Mullane}}]{Lindegren.etal:2016}
{Lindegren}, L., {Lammers}, U., {Bastian}, U., {et~al.} 2016, \aap, 595, A4

\bibitem[{{Martinez} {et~al.}(2015){Martinez}, {Stovall}, {Freire}, {Deneva},
  {Jenet}, {McLaughlin}, {Bagchi}, {Bates}, \& {Ridolfi}}]{Martinez2015}
{Martinez}, J.~G., {Stovall}, K., {Freire}, P.~C.~C., {et~al.} 2015, \apj, 812,
  143

\bibitem[{{Nelson} \& {Rappaport}(2003)}]{NelsonRapp2003}
{Nelson}, L.~A. \& {Rappaport}, S. 2003, \apj, 598, 431

\bibitem[{{{\"O}zel} \& {Freire}(2016)}]{Ozel2016}
{{\"O}zel}, F. \& {Freire}, P. 2016, \araa, 54, 401

\bibitem[{{{\"O}zel} {et~al.}(2012){{\"O}zel}, {Psaltis}, {Narayan}, \& {Santos
  Villarreal}}]{Ozel2012}
{{\"O}zel}, F., {Psaltis}, D., {Narayan}, R., \& {Santos Villarreal}, A. 2012,
  \apj, 757, 55

\bibitem[{{Papitto} {et~al.}(2013){Papitto}, {Ferrigno}, {Bozzo}, {Rea},
  {Pavan}, {Burderi}, {Burgay}, {Campana}, {di Salvo}, {Falanga},
  {Filipovi{\'c}}, {Freire}, {Hessels}, {Possenti}, {Ransom}, {Riggio},
  {Romano}, {Sarkissian}, {Stairs}, {Stella}, {Torres}, {Wieringa}, \&
  {Wong}}]{Papitto2013}
{Papitto}, A., {Ferrigno}, C., {Bozzo}, E., {et~al.} 2013, \nat, 501, 517

\bibitem[{{Patruno}(2012)}]{Patruno2012_b}
{Patruno}, A. 2012, \apjl, 753, L12

\bibitem[{{Patruno} \& {Watts}(2012)}]{Patruno_Watt2012}
{Patruno}, A. \& {Watts}, A.~L. 2012, ArXiv e-prints

\bibitem[{{Rappaport} {et~al.}(1982){Rappaport}, {Joss}, \&
  {Webbink}}]{1982ApJ...254..616R}
{Rappaport}, S., {Joss}, P.~C., \& {Webbink}, R.~F. 1982, \apj, 254, 616

\bibitem[{{Rix} \& {Bovy}(2013)}]{RixBovy2013}
{Rix}, H.-W. \& {Bovy}, J. 2013, \aapr, 21, 61

\bibitem[{{Sanna} {et~al.}(2016){Sanna}, {Papitto}, {Burderi}, {Bozzo},
  {Riggio}, {Di Salvo}, {Ferrigno}, {Rea}, \& {Iaria}}]{sanna2016}
{Sanna}, A., {Papitto}, A., {Burderi}, L., {et~al.} 2016, ArXiv e-prints

\bibitem[{{Verbunt}(1993)}]{Verbunt1993}
{Verbunt}, F. 1993, \araa, 31, 93

\bibitem[{{Wijnands} \& {van der Klis}(1998)}]{Wij1998}
{Wijnands}, R. \& {van der Klis}, M. 1998, \nat, 394, 344

\end{thebibliography}

\end{document}